\documentclass{ptephy_v1}

\preprintnumber{XXXX-XXXX}

\usepackage{url}
\usepackage{graphicx}
\usepackage{amsmath}
\usepackage{amssymb}
\usepackage{mathtools}
\usepackage{physics}
\usepackage{bm}
\usepackage{siunitx}
\usepackage{here}

\begin{document}

\title{Momentum reconstruction of charged particles \\ using multiple Coulomb scatterings \\ in a nuclear emulsion detector}

\author[1,*]{T.~Odagawa}
\affil{Department of Physics, Kyoto University, Kyoto 606-8502, Japan}

\author{Y.~Suzuki}
\affil{Graduate School of Science, Nagoya University, Nagoya 464-8602, Japan}

\author[2]{T.~Fukuda}

\author[1]{T.~Kikawa}

\author[2]{M.~Komatsu}

\author[1]{T.~Nakaya}

\author[2,3]{O.~Sato}
\affil{Kobayashi-Maskawa Institute for the Origin of Particles and the Universe, Nagoya University, Nagoya 464-8602, Japan}

\author[4]{H.~Shibuya}
\affil{Kanagawa University, Yokohama 221-8686, Japan\email{odagawa.takahiro.57w@st.kyoto-u.ac.jp}}

\author[1]{K.~Yasutome}

\begin{abstract}%
This paper describes a new method for momentum reconstruction of charged particles using multiple Coulomb scatterings in a nuclear emulsion detector with a layered structure of nuclear emulsion films and target materials.
The method utilizes the scattering angles of particles precisely measured in the emulsion films.
The method is based on the maximum likelihood to newly include information on the decrease of the energy as the particle travels through the detector.
According to the Monte Carlo simulations, this method can measure momentum with a resolution of 10\% for muons of $\SI{500}{\MeV}/c$ passing through the detector perpendicularly.
The momentum resolution is evaluated to be 10--20\%, depending on the momentum and emission angle of the particle.
By accounting for the effect of the energy decrease, the momentum can be reconstructed correctly with less bias, particularly in the low-momentum region.
We apply this method to measure the momentum of muon tracks detected in the NINJA experiment where the momentum is also measured independently by using the track range.
The two measurements agree well within experimental uncertainties, verifying the method experimentally.
This method will extend the measurable phase space of muons and hadrons in the NINJA experiment.
\end{abstract}

\subjectindex{C32, H16}

\maketitle

\section{Introduction\label{sec:introduction}}

In the current and future long-baseline neutrino oscillation experiments~\cite{T2K:2021xwb, NOvA:2021nfi, Abe:2018uyc, Acciarri:2015uup}, one of the major sources of systematic uncertainties comes from the uncertainty of the interaction models.
To understand the neutrino interactions for the reduction of the uncertainty, measurements of the kinematics of charged particles emitted from the interaction are important.
Since 2014, a series of experiments has been carried out, in Japan Proton Accelerator Research Complex (J-PARC), to measure neutrino-nucleus interactions in the sub- and multi-GeV neutrino energy region using nuclear emulsion detectors, so-called Neutrino Interaction research with Nuclear emulsion and J-PARC Accelerator (NINJA) experiment~\cite{Fukuda:2017clt, Hiramoto:2020gup, Oshima:2020ozn, NINJA:2022zbi}.

In this paper, we present a momentum reconstruction method of charged particles using a nuclear emulsion detector.
A charged particle traveling through a material is scattered by multiple Coulomb scatterings (MCS).
We call a difference of the measured angles before and after the material, ``scattering angle.''
The scattering angles are explained by the so-called Highland formula~\cite{Highland:1975pq, Lynch:1990sq} and are strongly correlated with the particle's momentum.
To measure the scattering angles, a detector with a high angular resolution is usually required because such angles are very small; e.g. a few \si{\milli\radian} for a $\SI{1}{\GeV}/c$ muon penetrating a \SI{500}{\um}-thick iron plate.
It is worth noting that the Highland formula is a relationship between the momentum and the statistical width of the collection of the scattering angles in one trajectory.
It follows that the detector should have the capability of measuring a large number of the scattering angles for a given particle along its track.
In particular, for hadrons, a large number of the scattering angles are required to be measured within a short distance, because hadrons sometimes become unmeasurable by absorption or charge exchange even before they fully lose the kinetic energy.
An Emulsion Cloud Chamber (ECC) is a nuclear emulsion detector, which satisfies these requirements.
ECC has an alternate structure of emulsion films and target materials.
As a charged particle travels across the detector, it is scattered by each target material and the scattering angles are measured by the nuclear emulsion films with a high angular resolution of a few mrad.
Making use of the high spatial resolution and high-sampling detector structure, the momentum reconstruction with MCS has been applied in the DONuT~\cite{Kodama:2007mw} and OPERA experiments~\cite{OPERA:2011aa}.

\begin{figure}[h]
    \centering
    \includegraphics[width = 0.6\textwidth]{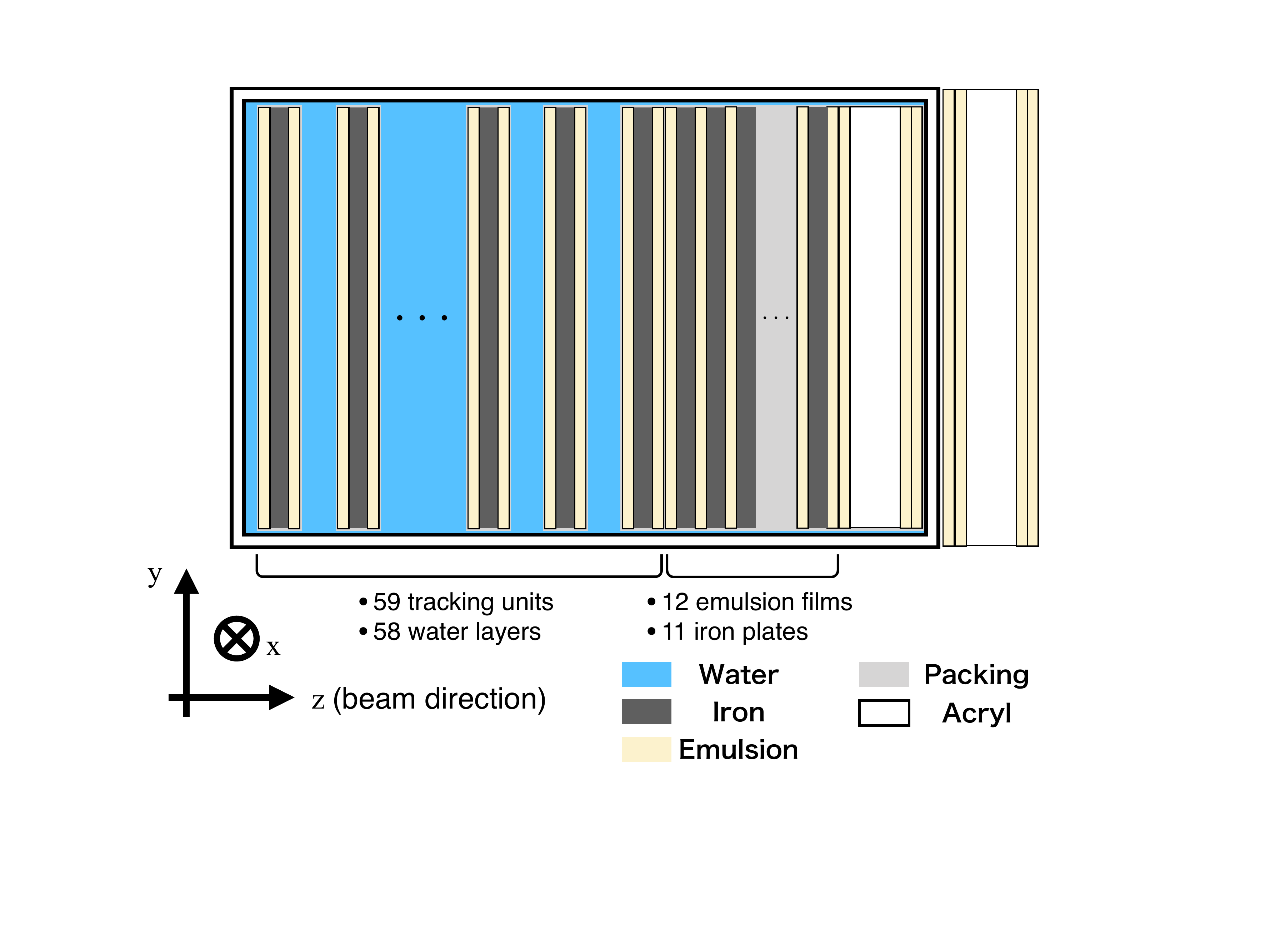}
    \caption[water ECC]{Water ECC used in the physics run of the NINJA experiment. Tracking units and \SI{2.3}{\mm}-thick water layers are alternately set perpendicularly to the neutrino beam direction. Each tracking unit consists of two emulsion films on both sides of a \SI{500}{\um}-thick iron plate.\label{fig:introduction:water_ecc}}
\end{figure}

To measure neutrino-water interactions, the NINJA experiment carried out its first physics run from November 2019 to February 2020.
The ECC used in the physics run consists of tracking units and \SI{2.3}{\mm}-thick water target layers as shown in Fig.~\ref{fig:introduction:water_ecc}.
Each tracking unit consists of two emulsion films on both sides of a \SI{500}{\um}-thick iron plate.
The tracking units and target water layers are placed perpendicularly to the neutrino beam direction.
There are 58 water layers and 70 iron plates in total in the detector.
Thus, the total thicknesses of water layers and iron plates in an ECC along the beam direction are \SI{133}{\mm} and \SI{35}{\mm}, respectively, and one ECC has about 2.5 radiation length units.

The energy region in the NINJA experiment is lower than that in the DONuT and OPERA experiments.
Thus, the charged particles from the neutrino interactions have lower momentum, from a few to several hundred $\si{MeV}/c$.
They lose a significant amount of their energy during their travel in material and sometimes stop inside the detector volume.
Using the track range in the detector or MCS, the water ECC can precisely measure the momentum of the charged particles.

In the NINJA experiment, the momentum reconstruction with MCS plays an important role because it does not require the particle tracks to be fully contained in the detector volume.
This is relevant in the sub- and multi-\si{\GeV} neutrino energy region, since most of the muons or charged pions from the neutrino interactions exit from the volume.
Therefore, the momentum reconstruction with a track range in the detector is not applicable to them, whereas that with MCS is actually essential.
So far, momentum reconstruction with MCS of charged particles has been applied in the NINJA experiment~\cite{Hiramoto:2020gup, Oshima:2020ozn, NINJA:2022zbi}.
However, there is still room for further improvement of its performance.
In the previous method, the momentum of particles is assumed to be unchanged during their travel in material.
In contrast, the new method considers the decrease of momentum due to the energy deposit of the charged particles inside the detector.
Besides, the scattering angles used in the analysis are also reconsidered in order to follow the Highland formula more appropriately.

The new method is based on a maximum likelihood constructed by each scattering angle following a Gaussian distribution.
The consideration of the energy deposit inside the momentum reconstruction with MCS is firstly proposed by and used in the MicroBooNE experiment~\cite{Abratenko:2017nki}, which is the neutrino interaction measurement with a liquid argon time-projection chamber (LArTPC).
In the method described in Ref.~\cite{Abratenko:2017nki}, each scattering angle is measured for liquid argon with the path length equivalent to the radiation length unit of liquid argon, since the LArTPCs typically have a uniform target material medium and a sufficient range of $\order{\si{\m}}$.
In the ECC, on the other hand, the method can be used for shorter tracks owing to the high angular resolutions and high-sampling detector structure.
The scattering angles by smaller path length can be still precisely measured with keeping the number of the measurements sufficient.
This method considers the energy deposit of charged particles inside the detector and only one parameter of the Highland formula is phenomenologically tuned using a particle gun Monte Carlo (MC) simulation.

The remaining of the paper is organized as follows.
In Sect.~\ref{sec:mcs}, MCS expressed by the Highland formula is explained.
The scattering angles used in the method, the construction of the likelihood, and the effects from the angular resolution and the energy deposit are shown in Sect.~\ref{sec:reconstruction}.
Section~\ref{sec:range_validation} describes the validation of the method using another method of momentum reconstruction with a track range in a muon range detector.
Future prospects of the method are discussed in Sect.~\ref{sec:prospects} and finally Sect.~\ref{sec:conclusions} concludes the paper.

\section{Multiple Coulomb scatterings in the water ECC\label{sec:mcs}}

When a charged particle travels across a medium, its direction changes due to MCS inside the medium.
The scattering angles are modeled by a Gaussian distribution with mean at zero and RMS, $\sigma_\mathrm{HL}(p\beta, w/X_0)$, expressed by the Highland formula;
\begin{equation}
    \sigma_\mathrm{HL}(p\beta, w/X_0) = \frac{13.6\,\mathrm{MeV}/c}{p\beta}|q|\sqrt{\frac{w}{X_0}}\left[1 + 0.038 \ln\left(\frac{wq^2}{X_0\beta^2}\right)\right].
    \label{eq:highland_formula_basic}
\end{equation}
Here, $p$ is the momentum, and $\beta$ is the velocity of the particle.
$q$ is the electric charge of the particle, and $w/X_0$ is the path length of the particle in a unit of the radiation length of the scattering medium.

In the water ECC, particles are mainly scattered by the \SI{500}{\um}-thick iron plates.
In this analysis, we only use the scatterings inside each tracking unit, although the scatterings in water layers are also potentially available.
The scatterings inside each tracking unit is mainly due to the iron plate with small contribution of the nuclear emulsion.
Table~\ref{tab:reconstruction:radiation_length} shows the radiation length and thickness of each material segment in the water ECC in the NINJA experiment\footnote{In the water ECC, SUS316L is used as ``iron''.}.
In the MC simulation, the values in Table~\ref{tab:reconstruction:radiation_length} are used.
A polystyrene sheet is used to support the emulsion and called ``base''.
The two emulsion films and one iron plate are vacuum-packed by an aluminum-laminated envelope sheet.
\begin{table}[t]
    \centering
    \caption[Radiation length of materials in the water ECC.]{Thickness and radiation length of each material in the water ECC. Base is a polystyrene sheet to support the emulsion.\label{tab:reconstruction:radiation_length}}
    \begin{tabular}{lccc} \hline
         Material & Thickness $t$ [mm] & Radiation length $X_0$ [mm] & $t/X_0$ [$\times 10^{-3}$]\\ \hline \hline
         Iron & 0.5 & 17.18 & 29.1 \\
         Water & 2.3 & 360.8 & 6.37\\
         Emulsion & 0.07 & 30.3 & 2.31\\
         Base (polystyrene) & 0.21 & 413.1 & 0.51\\ 
         Packing & 0.109 & 413.1 & 0.26 \\ \hline
    \end{tabular}
\end{table}
Using Eq.~\eqref{eq:highland_formula_basic} and values in Table~\ref{tab:reconstruction:radiation_length}, an RMS of the scattering angles is typically found at a few mrad for muons with $p\beta = \SI{1}{\GeV}/c$.
The angular resolution of the emulsion film ranges from sub to several mrad~\cite{Suzuki:2021euh}, and comparable to the typical scattering angles by the iron plate.

The Highland formula is an equation for the scattering angles projected to the plane parallel to the incident track direction.
Since the nuclear emulsion film is a three-dimensional tracking detector, two independent scattering angles can be calculated by projecting the scattered track to two orthogonal planes.
Details of the scattering angles used in the method are described in Sect.~\ref{sec:reconstruction}.

Equation~\eqref{eq:highland_formula_basic} indicates that the change of the transverse momentum of the particle is a stochastic event following the Gaussian distribution.
It follows that the momentum resolution and the accuracy of $\sigma_\mathrm{HL}$ would both improve by increasing the number of the measurements of the scattering angles.
Therefore, the method proposed here would show the best performance for the tracks coming from the upstream of the detector and penetrating the most downstream emulsion film, as the charged particles penetrate the largest number of the iron plates.

\section{Maximum likelihood-based method using MCS\label{sec:reconstruction}}

\subsection{Definition of the scattering angles\label{ssec:reconstruction:angle_definition}}

First, we introduce a right-handed Cartesian coordinate system to measure the scattering angles.
The $x$, $y$, and $z$ directions are defined as follows.
The $z$ direction is perpendicular to the surface of the emulsion films, which is almost consistent with the neutrino beam direction.
The $x$ and $y$ directions are the horizontal and vertical directions, respectively.
The measurement error of a film scanning system, called Hyper Track Selector (HTS)~\cite{Yoshimoto:2017ufm, Suzuki:2021euh}, along the $z$ direction is largely higher than that in the other two directions.
As mentioned in Sect.~\ref{sec:mcs}, the scattering angles are measured on two orthogonal planes, parallel to the incident track direction.
The planes are selected so that one of them is totally independent of the measurement error along the $z$ direction.

Three unit vectors $\hat{\bm{t}}$, $\hat{\bm{r'}}$, and $\hat{\bm{l'}}$ are defined as follows.
$\hat{\bm{t}}$ is parallel to the incident direction of the track and
\begin{align}
    \hat{\bm{l'}} &= \frac{\hat{\bm{z}} \times \hat{\bm{t}}}{|\hat{\bm{z}} \times \hat{\bm{t}}|} \\
    \hat{\bm{r'}} &= \hat{\bm{t}} \times \hat{\bm{l'}}.
\end{align}
Here, $\hat{\bm{z}}$ is the unit vector along the $z$ direction.
$\hat{\bm{l'}}$ is orthogonal to the $z$ direction and independent of its measurement error.
By definition, these three vectors constitute another Cartesian coordinate system.

The scattering angles of the particle are projected to the planes made by $\hat{\bm{r'}}$-$\hat{\bm{t}}$ and $\hat{\bm{l'}}$-$\hat{\bm{t}}$ vectors.
Each angle of the particle in a nuclear emulsion film is measured as a vector form of the tangent; $\bm{a} = (\tan\theta_x, \tan\theta_y, 1)$ by HTS~\cite{Yoshimoto:2017ufm, Suzuki:2021euh}.
Using the track directions before and after the scattering, the scattering angles on each plane can be calculated as follows:
\begin{align}
    \begin{split}
        \Delta\theta_\mathrm{rad'} &= \arctan\left(\frac{\bm{a}_1\cdot \hat{\bm{r'}}}{\bm{a}_1 \cdot \hat{\bm{t}}} \right) \\
        &= \!\begin{multlined}[t]
        \arctan\left( \left. \frac{-\tan\theta_{x0}\tan\theta_{x1} - \tan\theta_{y0}\tan\theta_{y1} + \tan^2\theta_{x0} + \tan^2\theta_{y0}}{\sqrt{\tan^2\theta_{x0} + \tan^2\theta_{y0} + (\tan^2\theta_{x0} + \tan^2\theta_{y0})^2}} \right. \right. \\
        \hspace{0.3\textwidth} \left. \left. \middle/ \frac{\tan\theta_{x0}\tan\theta_{x1} + \tan\theta_{y0}\tan\theta_{y1} + 1}{\sqrt{\tan^2\theta_{x0} + \tan^2\theta_{y0} + 1}} \right. \right)
        \end{multlined}
    \end{split} \label{eq:mcs_radial_angle_difference} \\
    \begin{split}
         \Delta\theta_\mathrm{lat'} &= \arctan\left(\frac{\bm{a}_1 \cdot \hat{\bm{l'}}}{\bm{a}_1 \cdot \hat{\bm{t}}} \right) \\
         &= \!\begin{multlined}[t]
         \arctan\left(\left. \frac{-\tan\theta_{y0}\tan\theta_{x1} + \tan\theta_{x0}\tan\theta_{y1}}{\sqrt{\tan^2\theta_{x0} + \tan^2\theta_{y0}}} \right. \right. \\
         \hspace{0.3\textwidth} \left. \left. \middle/ \frac{\tan\theta_{x0}\tan\theta_{x1} + \tan\theta_{y0}\tan\theta_{y1} + 1}{\sqrt{\tan^2\theta_{x0} + \tan^2\theta_{y0} + 1}} \right. \right).
         \end{multlined}
    \end{split} \label{eq:mcs_lateral_angle_difference}
\end{align}
Here, $\bm{a}_i = (\tan\theta_{xi}, \tan\theta_{yi}, 1), (i = 0, 1)$ are the direction vectors of the track before and after the scattering, respectively, and $\hat{\bm{t}} = \bm{a_0}/|\bm{a_0}|$.
By definition, $\Delta\theta_\mathrm{rad'}$ and $\Delta\theta_\mathrm{lat'}$ are scattering angles on the $\hat{\bm{r'}}$-$\hat{\bm{t}}$ and $\hat{\bm{l'}}$-$\hat{\bm{t}}$ planes, respectively.
The track before the scattering is always along the $\hat{\bm{t}}$ direction and the direction of the track after the scattering projected to each plane is treated as the scattering angle.

\begin{figure}[t]
    \centering
    \includegraphics[width = 0.7\textwidth]{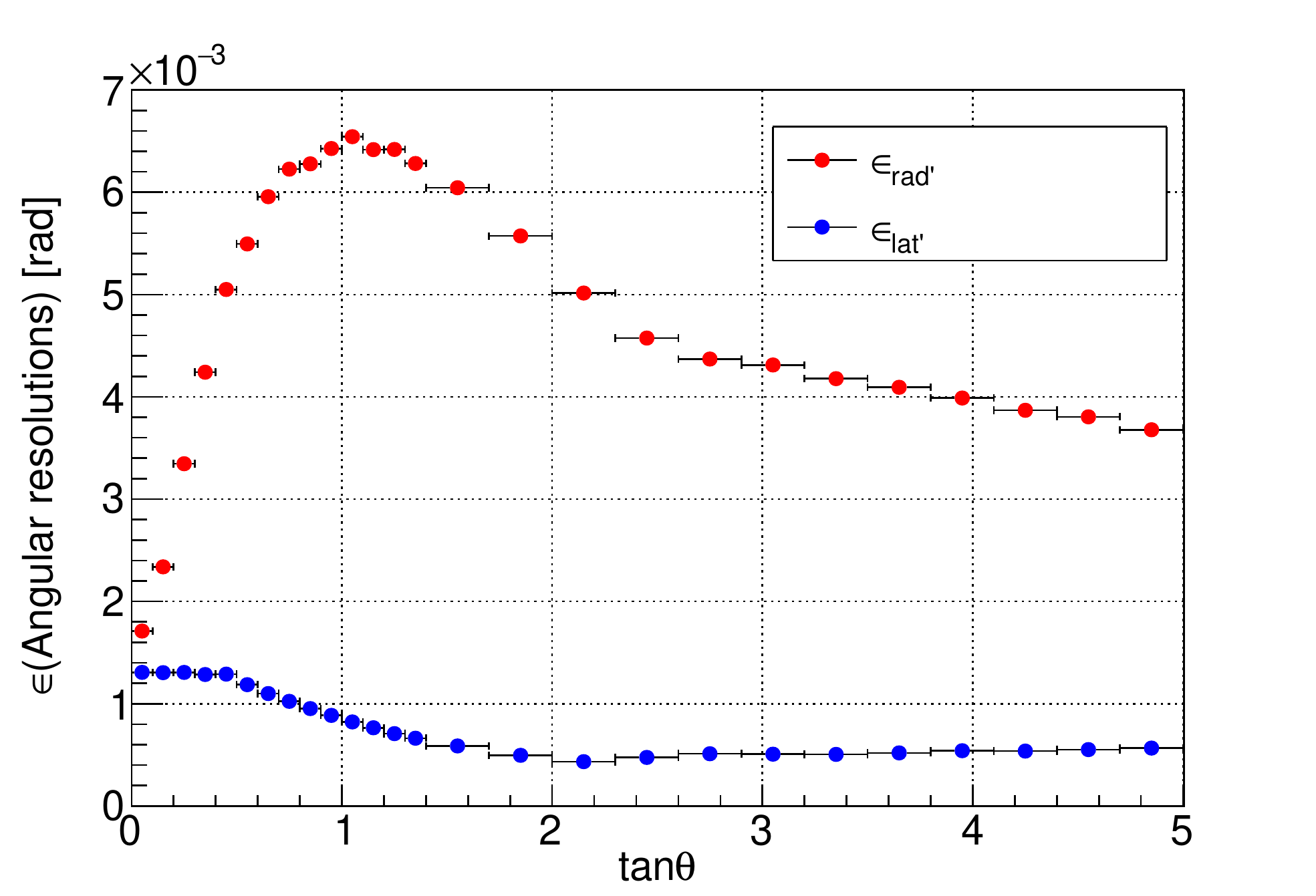}
    \caption[New radial'/lateral' angle precision]{Angular resolutions of the detector for each scattering angle. The red plots are the resolution of $\Delta\theta_\mathrm{rad'}$ and blue ones are of $\Delta\theta_\mathrm{lat'}$~\cite{Suzuki:2021euh}. $\epsilon_\mathrm{rad'}$ and $\epsilon_\mathrm{lat'}$ become the same at $\tan\theta = 0$, which is verified by extrapolating the plots.\label{fig:new_rad_lat_precision}}
\end{figure}

The scattering angles on both planes can be treated as completely identical, except for the angular resolutions of the detector, which are shown in Fig.~\ref{fig:new_rad_lat_precision}~\cite{Suzuki:2021euh} for each scattering angle.
As introduced above, the angular resolution of $\Delta\theta_\mathrm{lat'}$ is much better than that of $\Delta\theta_\mathrm{rad'}$ in all angle regions since the $\hat{\bm{l'}}$ direction is independent of the measurement error along the $z$ direction.
The angular resolution of $\Delta\theta_\mathrm{rad'}$ is degraded with a larger angle in the $\tan\theta < 1$ region, while it gets improved in the region of $\tan\theta > 1$.
Hereafter, $\tan\theta = \sqrt{\tan^{2}\theta_{x0} + \tan^{2}\theta_{y0}}$ represents the tangent of the track angle before each scattering with respect to the direction perpendicular to the film surface.
This is because the measurement error along the $z$ direction worsen the resolution with the larger angle, whereas the longer baseline improves it\footnote{The angular resolution of the scattering angles, $\sigma_\theta$, is related to that of $\tan\theta$, $\sigma_{\tan\theta}$, as $\sigma_{\tan\theta} \simeq \sigma_{\theta} / \cos^2\theta$. Thus, larger $\tan\theta$, i.e. longer baseline, makes $\sigma_\theta$ better even if $\sigma_{\tan\theta}$ is the same. In this study, $\sigma_\theta$ corresponds to $\epsilon(\tan\theta)$.}.
These two effects are comparable when $\tan\theta \simeq 1$.

\subsection{Maximum likelihood implementation\label{ssec:reconstruction:likelihood}}

The two scattering angles, $\Delta\theta_\mathrm{rad'}$ and $\Delta\theta_\mathrm{lat'}$, follow the Gaussian probability function:
\begin{equation}
    \begin{split}
    f(\Delta\theta; p\beta, w/X_0, \tan\theta) &= (2\pi\sigma(p\beta, w/X_0, \tan\theta)^2)^{-1/2} \\
    & \quad \times \exp\left[-\frac{1}{2}\left(\frac{\Delta\theta}{\sigma(p\beta,w/X_0, \tan\theta)}\right)^2\right].
    \label{eq:prob_func_one_angle_difference_true}
    \end{split}
\end{equation}
The track segment reconstructed in the emulsion film has a positional information on the upstream surface of the base.
$w/X_0$ is proportional to the distance between positions of the two track segments before and after the scattering.
$\Delta\theta$ represents $\Delta\theta_\mathrm{rad'}$ or $\Delta\theta_\mathrm{lat'}$, and $\sigma(p\beta,w/X_0, \tan\theta)$ includes the MCS part, $\sigma_\mathrm{HL}$, and the angular resolution part, $\epsilon_\mathrm{rad'}(\tan\theta)$ or $\epsilon_\mathrm{lat'}(\tan\theta)$ shown in Fig.~\ref{fig:new_rad_lat_precision}.
\begin{equation}
    \sigma(p\beta,w/X_0, \tan\theta) = \sqrt{\left(\sigma_\mathrm{HL}(p\beta,w/X_0)\right)^2 + \left(\epsilon(\tan\theta)\right)^2}.
    \label{eq:sigma_hl_ang_res}
\end{equation}

To obtain the likelihood, the product of the probability functions (Eq.~\eqref{eq:prob_func_one_angle_difference_true}) is considered.
When the particle penetrates $N$ iron plates, the scattering angles $\Delta\theta_i\,(i = 0, 1, \dots N-1)$ can be calculated across each iron plate.
Thus, the likelihood is written as follows:
\begin{equation}
   \begin{split}
        &L ((\bm{\Delta\theta}); (\bm{p\beta}), (\bm{w/X_0}), (\bm{{\tan\theta}})) \\
        &= (2\pi)^{-N} \times \prod_{j = 0}^{N-1}(\sigma_\mathrm{rad'}((p\beta)_j, w/X_0)_j, \tan\theta_j))^{-1} \times \prod_{j = 0}^{N-1}(\sigma_\mathrm{lat'}((p\beta)_j, w/X_0)_j, \tan\theta_j))^{-1} \\
        &\times \exp\left[-\frac{1}{2}\sum_{j = 0}^{N-1}\left(\left(\frac{\Delta\theta_{\mathrm{rad'}, j}}{\sigma_\mathrm{rad'}((p\beta)_j, (w/X_0)_j, \tan\theta_j)}\right)^2 + \left(\frac{\Delta\theta_{\mathrm{lat'}, j}}{\sigma_\mathrm{lat'}((p\beta)_j, (w/X_0)_j, \tan\theta_j)}\right)^2 \right)\right].
        \label{eq:mcs_likelihood}
    \end{split} 
\end{equation}
Here, $\sigma_\mathrm{rad'}$ and $\sigma_\mathrm{lat'}$ are expressed as Eq.~\eqref{eq:sigma_hl_ang_res} with $\epsilon = \epsilon_\mathrm{rad'}$ and $\epsilon_\mathrm{lat'}$, respectively.
$(\bm{\Delta\theta}) = (\Delta\theta_{\mathrm{rad'},0}, \Delta\theta_{\mathrm{rad'},1}, \dots \Delta\theta_{\mathrm{rad'},N-1}, \Delta\theta_{\mathrm{lat'},0}, \Delta\theta_{\mathrm{lat'},1}, \dots \Delta\theta_{\mathrm{lat'},N-1})$, $(\bm{w/X_0}) = ((w/X_0)_0, (w/X_0)_1, \dots (w/X_0)_{N-1})$, and $(\bm{{\tan\theta}}) = (\tan\theta_0, \tan\theta_1, \dots \tan\theta_{N-1})$ are the scattering angles, path lengths in the radiation length unit, and track directions before each scattering, respectively.
The momentum of the charged particle penetrating the detector volume also decreases, thus $p\beta$ is changed at each iron plate as $(\bm{p\beta}) = ((p\beta)_0, (p\beta)_1, \dots (p\beta)_{N-1})$, where $(p\beta)_0$ is our intended goal.

\begin{figure}[b]
    \centering
    \includegraphics[width = 0.7\textwidth]{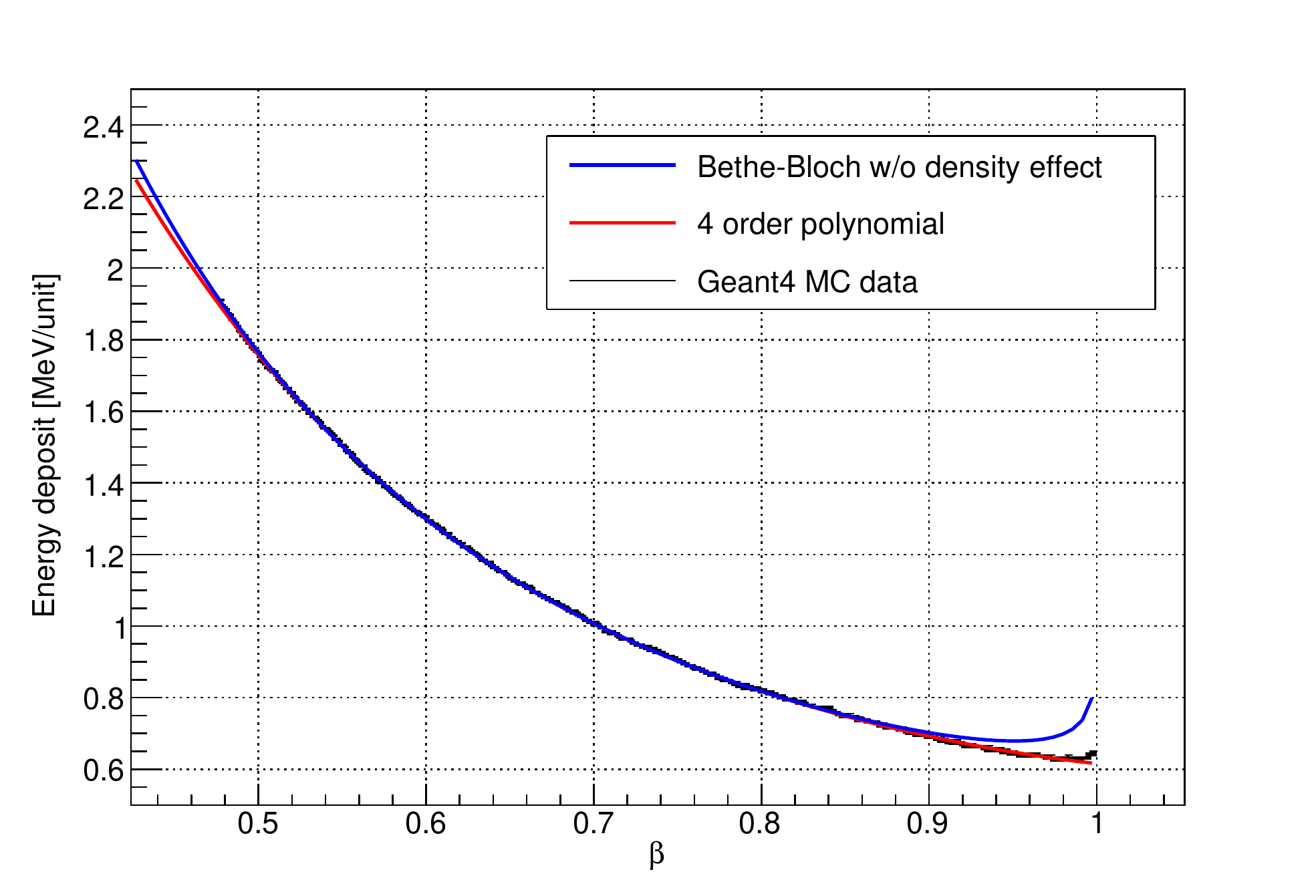}
    \caption[iron energy deposit]{The energy deposit across one iron plate of muon perpendicularly penetrating the emulsion films calculated by MC-truth information. The plot is fitted by the polynomial function and Bethe Bloch function without density effect as a reference. Without density effect, the Bethe Bloch function results in higher value than the expected one in the region of $\beta > 0.9$.\label{fig:iron_bb_pol4_compare}}
\end{figure}

The decrease of $p\beta$ between each pair of adjacent iron plates is treated as follows.
The MC simulation of the detector, based on \textsc{Geant4}~\cite{Agostinelli:2002hh}, is developed.
Muon particle guns with different $\beta$ ranging between 0.5 and 0.999 ($\gamma \simeq 22$) are generated perpendicularly to the film surface in the water ECC.
The energy deposit in one iron plate and one water layer is calculated from the MC-truth information.
The averaged energy deposit of each $\beta$ is plotted and fitted by a fourth order polynomial function as shown in Fig.~\ref{fig:iron_bb_pol4_compare}.
When the initial value of $p\beta$, $(p\beta)_0$, is set, the decrease of $p\beta$ is calculated one by one, assuming that the particle is a muon using Eqs.~\eqref{eq:pbeta_to_energy} and \eqref{eq:energy_deposition}.
\begin{equation}
    (p\beta)_i = \frac{p_i^2}{E_i} = \frac{E_i^2 - M^2}{E_i}
    \label{eq:pbeta_to_energy}
\end{equation}
\begin{equation}
    E_{i+1} = E_i - \Delta E_i.
    \label{eq:energy_deposition}
\end{equation}
Here, $M$ is the muon mass, $p_i$ and $E_i$ are the momentum and energy of the particle at the $i$-th iron plate, respectively.
$\Delta E_i$ is calculated using $\beta$, the energy deposit in Fig.~\ref{fig:iron_bb_pol4_compare} and the path length in an iron plate the particle penetrates, and the similar values for a water layer.
This process is repeated until the most downstream iron plate is reached or the kinetic energy of the particle becomes negative.

According to Eq.~\eqref{eq:mcs_likelihood} and the treatment of the decrease of $p\beta$, the likelihood is dependent only on the positions and angles measured by the emulsion films and has one parameter, $(p\beta)_0$.
$(p\beta)_0$ which maximizes the likelihood in Eq.~\eqref{eq:mcs_likelihood} corresponds to the reconstructed momentum of the particle.
In this likelihood, when $\Delta\theta_i$ is larger than three times the RMS of them, such scattering angles are not used but only the energy deposit is considered, since the Gaussian approximation is only applicable to the central 98\% of the scattering angles.
For example, in the MC simulation of $\SI{500}{\MeV}/c$ muon particle guns perpendicular to the film surface, 2\% of the scattering angles are discarded by this cut.

Instead of maximizing the likelihood in Eq.~\eqref{eq:mcs_likelihood}, the negative log likelihood is minimized using the MINUIT framework~\cite{James:1975dr} in ROOT.
\begin{equation}
    \begin{split}
        -l &= -2\log(L) + N\log(2\pi) \\
        &= 2 \sum_{j = 0}^{N-1} \left[ \log(\sigma_\mathrm{rad'}((p\beta)_j, (w/X_0)_j, \tan\theta_j)) + \log(\sigma_\mathrm{lat'}((p\beta)_j, (w/X_0)_j, \tan\theta_j))\right] \\
        & \quad + \sum_{j = 0}^{N-1} \left[\left(\frac{\Delta\theta_{\mathrm{rad'}, j}}{\sigma_\mathrm{rad'}((p\beta)_j, (w/X_0)_j, \tan\theta_j)}\right)^2 + \left(\frac{\Delta\theta_{\mathrm{lat'}, j}}{\sigma_\mathrm{lat'}((p\beta)_j, (w/X_0)_j, \tan\theta_j)}\right)^2 \right].
        \label{eq:mcs_negative_log_likelihood}
    \end{split}
\end{equation}
Here, a constant term $-N\log(2\pi)$ is subtracted from $-2\log(L)$.

\subsection{Parameter tuning\label{ssec:reconstruction:tuning}}

Instead of using Eq.~\eqref{eq:highland_formula_basic}, the Highland formula is phenomenologically tuned using the detector MC simulation.
The coefficients in the Highland formula (i.e. 13.6 and 0.038) are determined in a way that renders the formula globally applicable for various materials.
Each experiment tunes the parameters depending on their detector structures or scattering medium~\cite{Kodama:2007mw, OPERA:2011aa, Abratenko:2017nki}.
In this method, the coefficient $\SI{13.6}{\MeV}/c$ in Eq.~\eqref{eq:highland_formula_basic} is replaced by a free parameter $S$.
\begin{equation}
    \sigma_\mathrm{HL}(p\beta, w/X_0) = \frac{S}{p\beta}\sqrt{\frac{w}{X_0}}\left[1 + 0.038 \ln\left(\frac{w}{X_0}\right)\right].
    \label{eq:highland_formula_tune_parameter}
\end{equation}
Hereafter, $q$ is always set to unity since we only consider muons, protons, or charged pions.
In addition, $\beta$ in the logarithm function is also set to one because the muon in this simulation is relativistic ($\beta > 0.93$).
Tuning the other coefficient (i.e. 0.038) has a smaller effect compared to that of $S$.
Therefore, only the tuning of $S$ is considered in this method.

\begin{figure}[h]
    \centering
    \includegraphics[width = 0.7\textwidth]{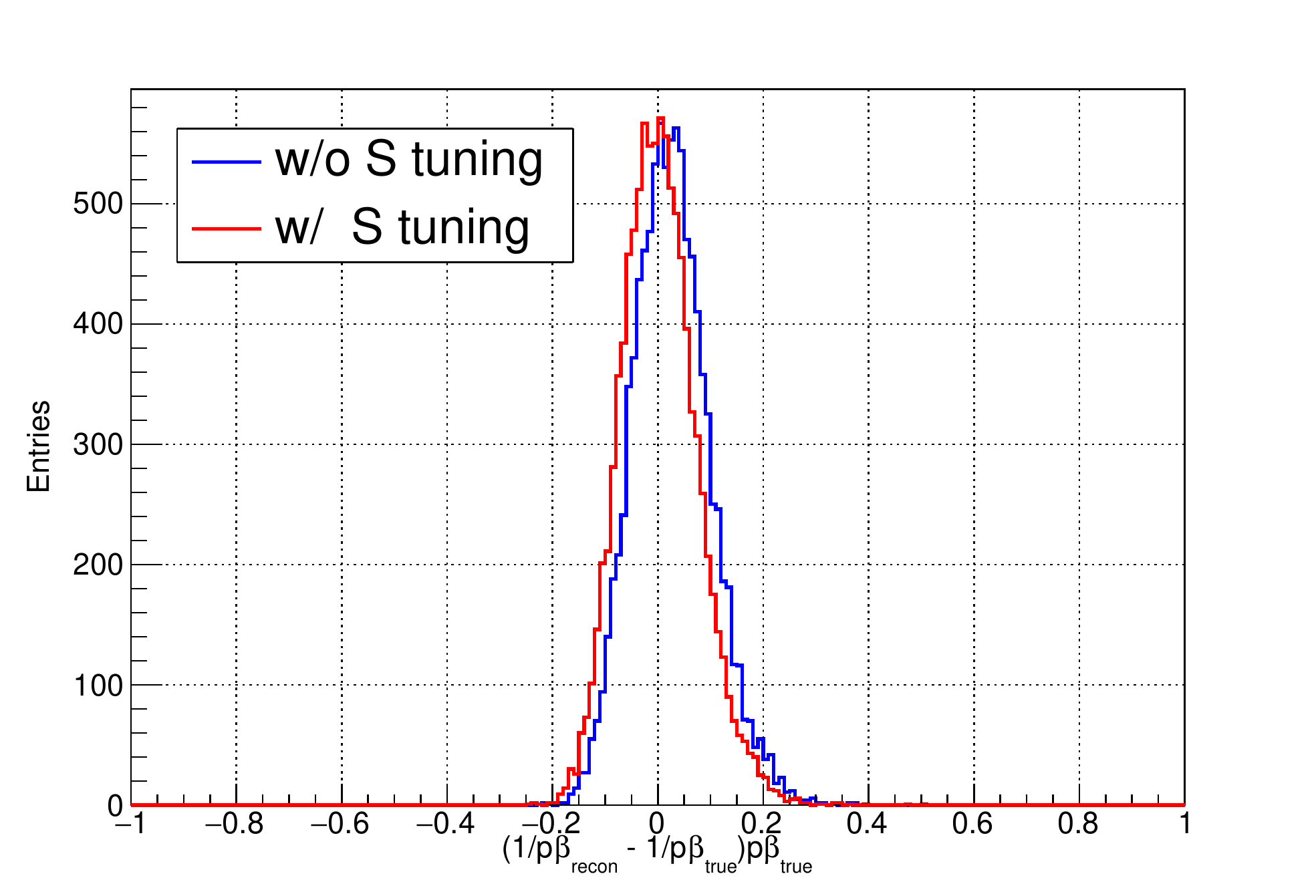}
    \caption{Relative residual distribution of $1/p\beta$ of $p = \SI{500}{\MeV}/c$ muon particle guns before (blue) and after (red) the $S$ parameter tuning. Ten thousand muons are simulated for each distribution.\label{fig:s_tune_hist_compare}}
\end{figure}

\begin{figure}[H]
    \centering
    \includegraphics[width = 0.7\textwidth]{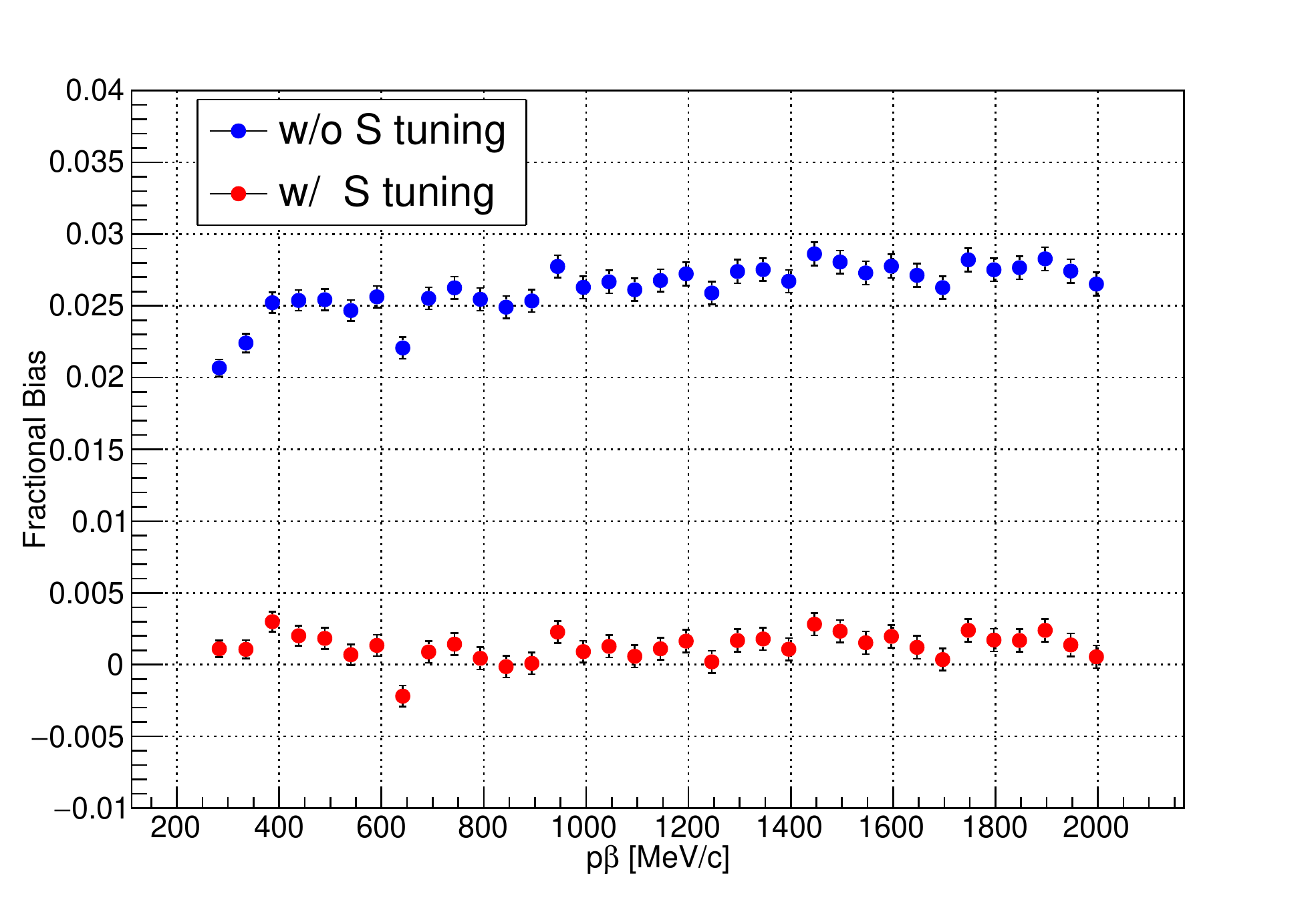}
    \caption{Fractional bias of the relative $1/p\beta$ as a function of $p\beta$ before (blue) and after (red) the $S$ parameter tuning. The bias is reduced from 2.7\% to around 0.1\%. The vertical error bars are the statistical ones in the Gaussian fitting.\label{fig:s_tune_compare}}
\end{figure}

To consider only the Highland formula, the angular resolution is set to zero in this tuning.
First, the momentum of muon particle guns is reconstructed with $S = \SI{13.6}{\MeV}/c$.
The negative log likelihood with $S = \SI{13.6}{\MeV}/c$ is minimized for each particle and the reconstructed values are obtained.
This reconstruction results in a few percent bias in the $1/p\beta$ relative residual distribution.
Figure~\ref{fig:s_tune_hist_compare} shows the relative residual distributions of $1/p\beta$ for muon particle guns of $p = \SI{500}{\MeV}/c$.
The blue histogram in Fig.~\ref{fig:s_tune_hist_compare} shows the distribution before the tuning and there is a few \% bias.
The bias is obtained for the muon particle guns with $p = 300\text{--}\SI{2000}{\MeV}/c$.
The mean of the fractional biases of each residual distribution is obtained as 2.7\% in this momentum region.
Figure~\ref{fig:s_tune_compare} shows the biases of particle gun simulations with different momenta before and after the tuning.

When $1/p\beta$ has a bias $\delta$, the $1/p\beta$ reconstruction should be scaled by $1/(1 + \delta)$.
The measured scattering angles are not changed, thus the parameter $S$ should be scaled as
\begin{equation}
    \sigma_\mathrm{HL} \propto \frac{\SI{13.6}{\MeV}/c}{p\beta} \to \sigma_\mathrm{HL} \propto \frac{S}{(1 + \delta) \times p\beta}
\end{equation}
to adjust the reconstructed $1/p\beta$.
Then, $S$ is tuned as $S = 1.027 \times \SI{13.6}{\MeV}/c \sim \SI{14.0}{\MeV}/c$ in this study and the bias is expected to be reduced.

After the parameter tuning, the momentum reconstruction with MCS is performed again and the distribution is changed, as shown on the red histogram in Fig.~\ref{fig:s_tune_hist_compare}.
The bias is now around 0.1\%.
This bias after the tuning for different momenta are shown on the red plots in Fig.~\ref{fig:s_tune_compare}.
The biases are almost similar in this momentum region, and the tuning is independent of the momentum.

\subsection{Effect from the angular resolution\label{ssec:reconstruction:effect_precision}}

The effect on the performance of the momentum reconstruction from the angular resolution is also evaluated using the MC simulation.
First, the particle gun MC simulation is performed using the MC-truth information.
The muon particle gun is generated in the most upstream water layer of the water ECC and directed downstream perpendicularly to the film surface.
According to the detector structure, the muon penetrates at most 69 iron plates until it escapes from the detector volume.
When the angular resolution is set to zero in the likelihood, the momentum resolution is typically 7\% with 0.1\% bias.
Hereafter, we refer to the width of the relative residual distribution of $1/p\beta$ as ``the momentum resolution''.
Then, the MC simulation and reconstruction are re-performed with the smearing and angular resolution.

\begin{figure}[h]
    \centering
    \includegraphics[width = 0.7\textwidth]{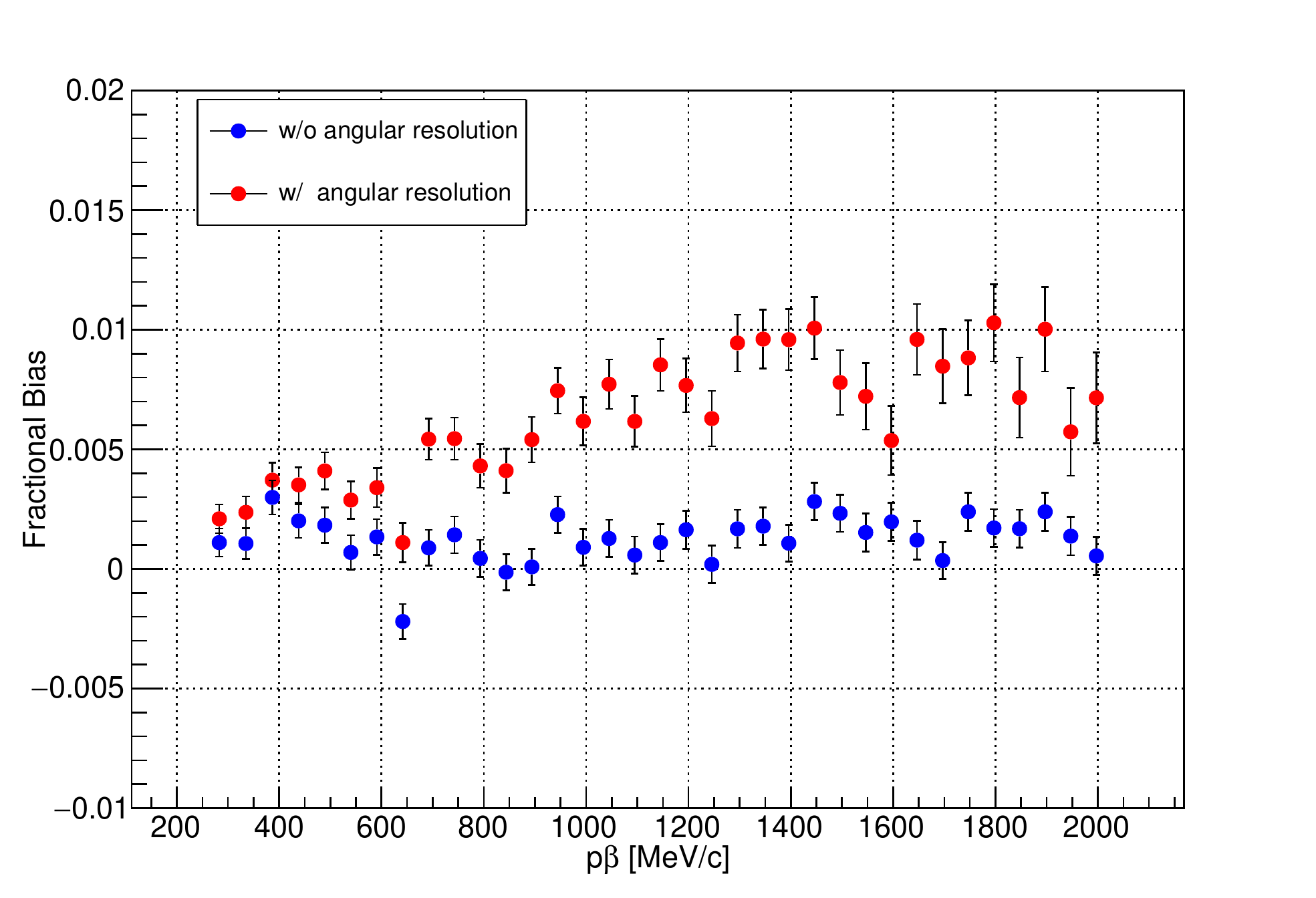}
    \caption{Fractional bias of the relative residual distributions of the $1/p\beta$ as a function of $p\beta$. The blue plots show the performance in the MC-truth level simulation, while the red ones consider the angular resolution. The vertical error bars are the statistical ones in the Gaussian fitting.\label{fig:bias_angres_mom_compare}}
\end{figure}

\begin{figure}[h]
    \centering
    \includegraphics[width = 0.7\textwidth]{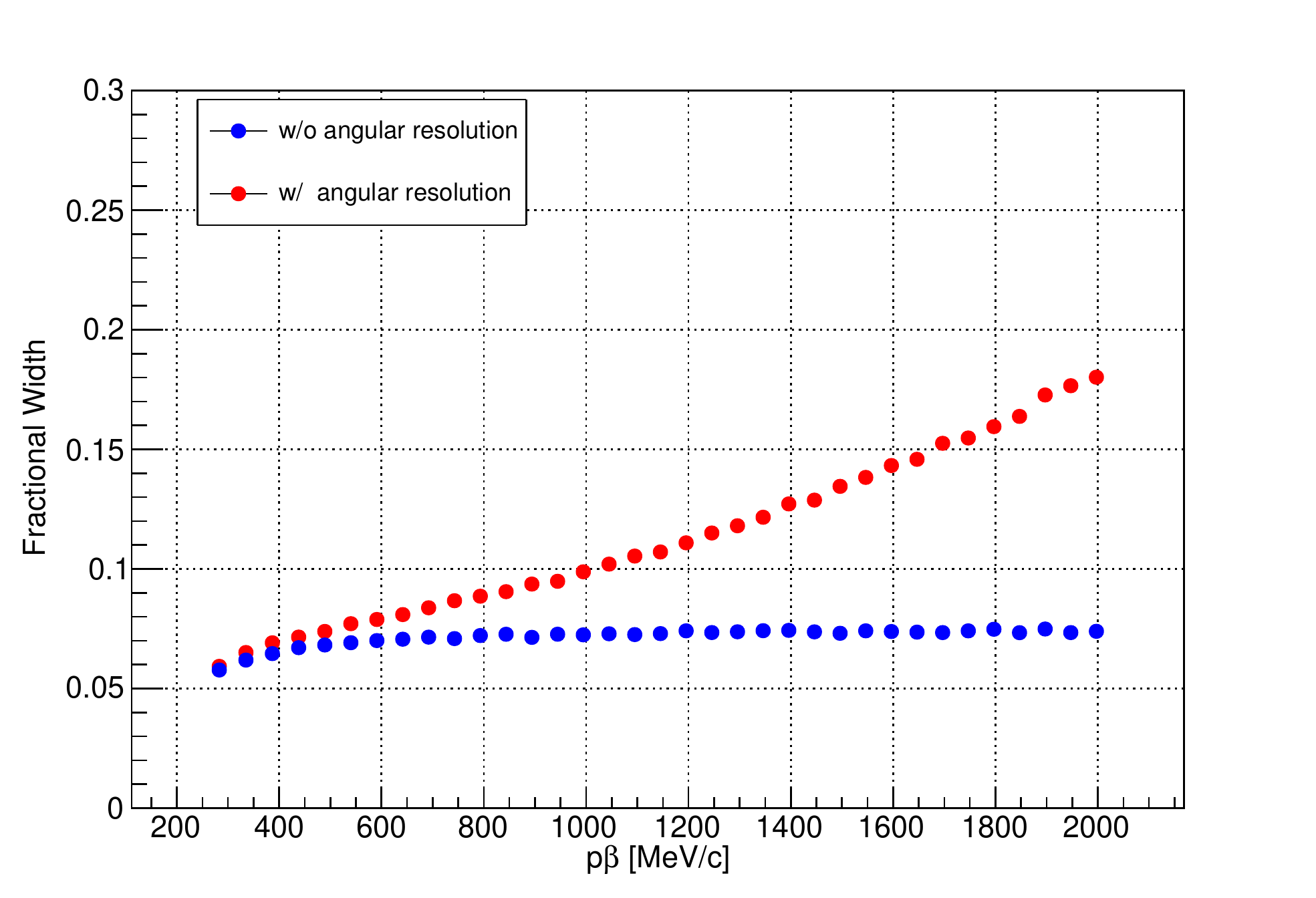}
    \caption{Fractional width of the relative residual distributions of the $1/p\beta$ as a function of $p\beta$. The blue plots show the performance in the MC-truth level simulation, while the red ones consider the angular resolution. The vertical errors are negligible.\label{fig:width_angres_mom_compare}}
\end{figure}

When the angular resolution is considered, the performance as a function of the momentum is changed from the blue to the red plots in Figs.~\ref{fig:bias_angres_mom_compare} and \ref{fig:width_angres_mom_compare}.
The performance is determined by the ratio of the magnitude of MCS to the angular resolution.
For the higher momentum, the magnitude of MCS decreases while the resolution remains unchanged.
Thus, the likelihood is primarily determined by the angular resolution and the momentum information is harder to obtain from the likelihood.
In the high-momentum region, larger fractional bias is seen, i.e. the reconstructed momentum value is smaller than the true one.
This is because, when the magnitude of MCS is much smaller than the angular resolution, the reconstructed momentum is biased to the lower value.

\begin{figure}[H]
    \centering
    \includegraphics[width = 0.7\textwidth]{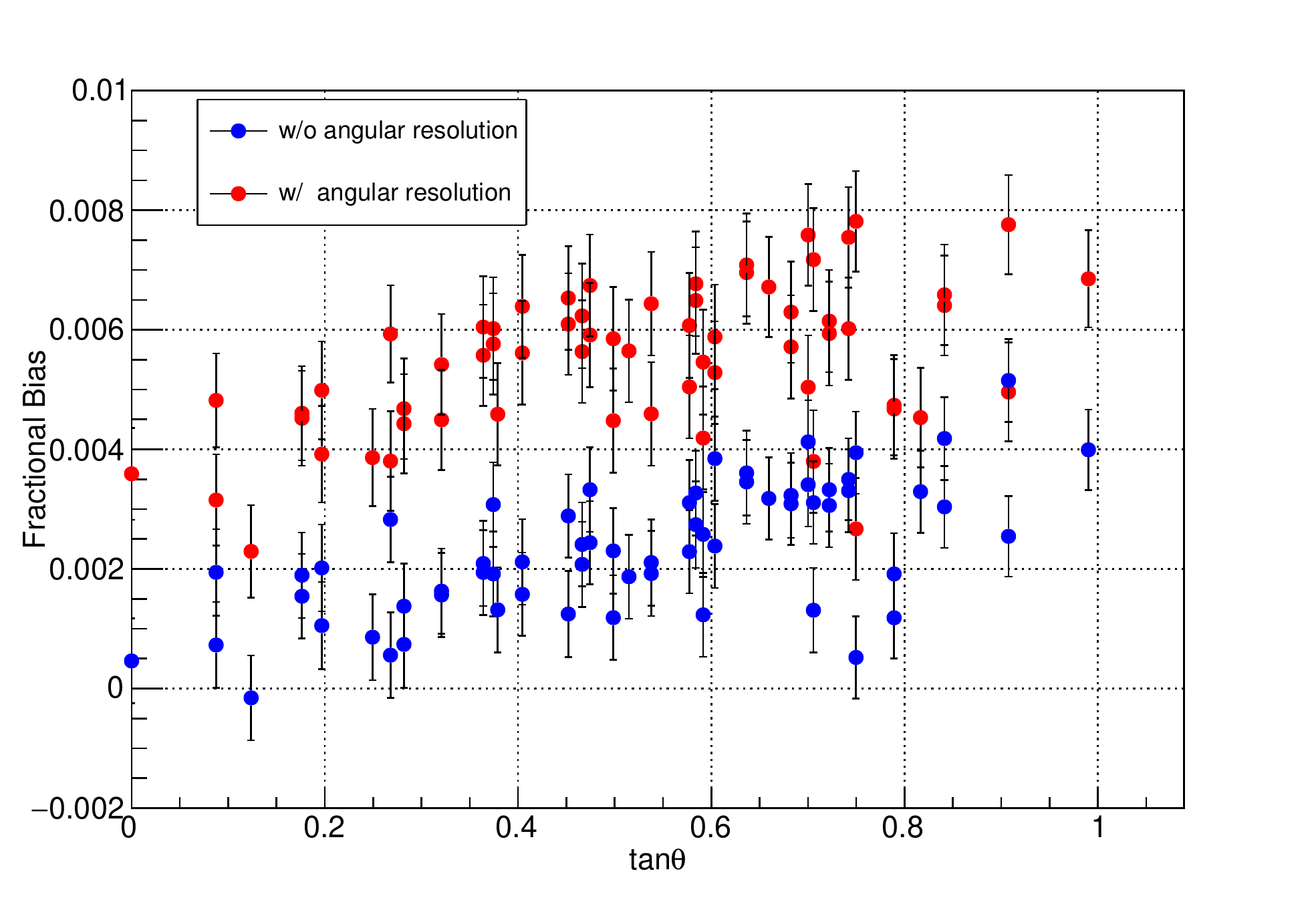}
    \caption{Fractional bias of the relative residual distributions of the $1/p\beta$ as a function of the incident angle. The blue plots show the performance in the MC-truth level simulation, while the red ones consider the angular resolution. The vertical error bars are the statistical ones in the Gaussian fitting.\label{fig:bias_angres_ang_compare}}
\end{figure}

\begin{figure}[H]
    \centering
    \includegraphics[width = 0.7\textwidth]{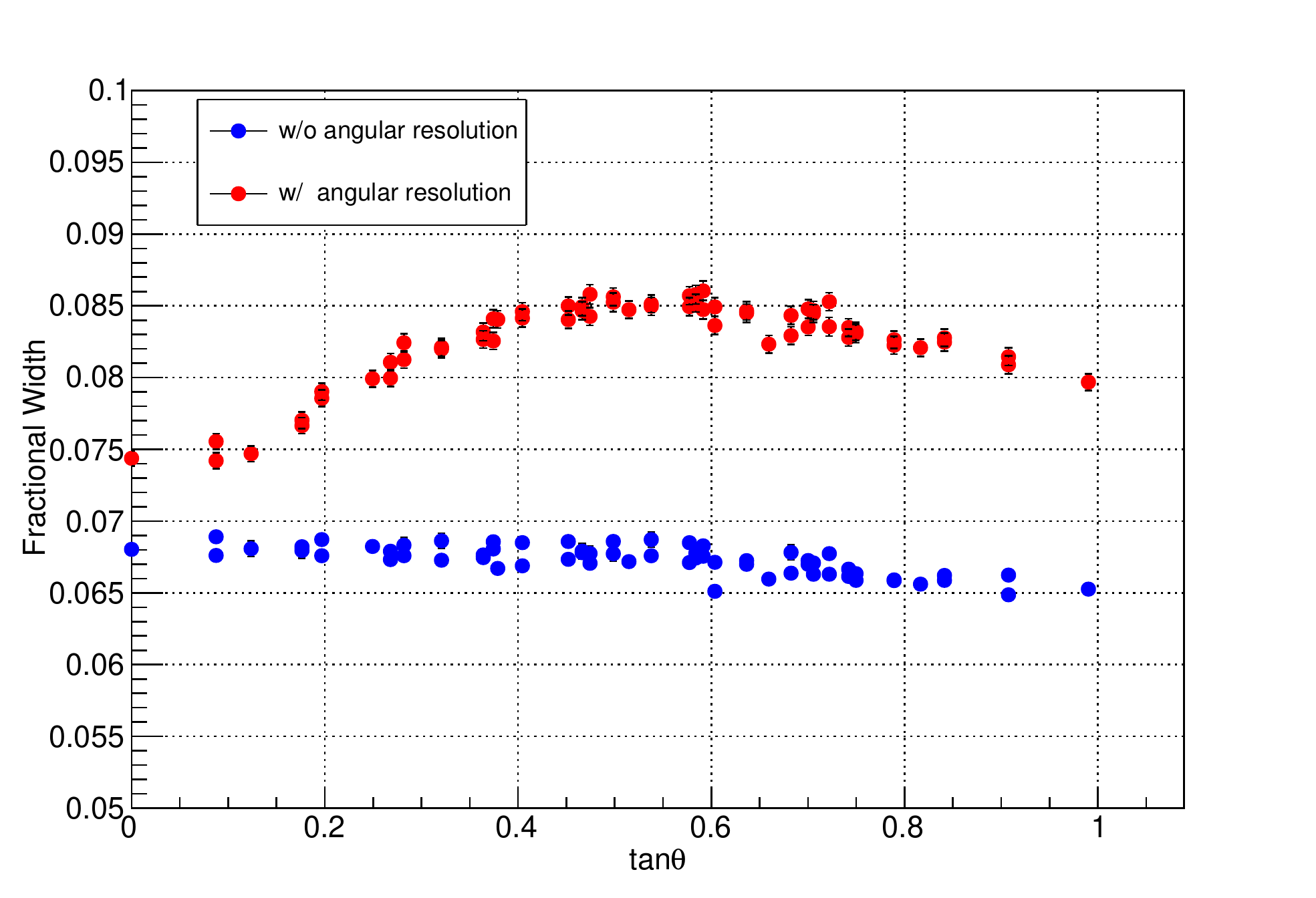}
    \caption{Fractional width of the relative residual distributions of the $1/p\beta$ as a function of the incident angle. The blue plots show the performance in the MC-truth level simulation, while the red ones consider the angular resolution. The vertical error bars are the statistical ones in the Gaussian fitting.\label{fig:width_angres_ang_compare}}
\end{figure}

The fractional bias and width as a function of the incident angle of the particle are similarly studied using the particle gun MC simulation.
In this study, muons with $p = \SI{500}{\MeV}/c$ are generated with different incident angles.
When the angular resolutions are considered, the performance is changed from the blue plots to the red ones in Figs.~\ref{fig:bias_angres_ang_compare} and \ref{fig:width_angres_ang_compare}.
Here, the muon particle guns are required to penetrate the most downstream emulsion film of the ECC, i.e. muon penetrates 69 iron plates.
According to the angular resolution in Fig.~\ref{fig:new_rad_lat_precision}, the performance is expected to be the worst around $\tan\theta = 1$ and to improve in the larger angle region\footnote{In this section, $\tan\theta$ represents the incident angle of the muon particle gun with respect to the direction perpendicular to the film surface.}.
In addition to the angular resolution, the MCS also increases as the path length becomes longer.
Thus, the momentum resolution improves with a larger angle for $\tan\theta > 0.6$.

Considering the effect from the angular resolutions of the nuclear emulsion film, the momentum resolution is 10--20\% in a region of momentum from a few hundred $\si{\MeV}/c$ to $\SI{2}{\GeV}/c$ and angle up to $\tan\theta = 1.0$.
The bias is less than 1\% up to $p\beta = \SI{1.5}{\GeV}/c$, which is acceptable for the interest region of the NINJA experiment, i.e. a few to several hundred $\si{MeV}/c$.

A mis-modeling of the angular resolution leads to the additional bias.
According to Eq.~\eqref{eq:sigma_hl_ang_res}, the mis-modeling of the angular resolution results a wrong evaluation of $\sigma_\mathrm{HL}$: e.g. the smaller angular resolution leads to the larger $\sigma_\mathrm{HL}$, thus the smaller reconstructed value of $p\beta$.
In the low-momentum region, the mis-modeling does not have a significant effect since $\sigma_\mathrm{HL}$ is sufficiently larger than the angular resolution.
On the other hand, the upper limit of the applicable momentum region is more sensitive to this mis-modeling.

\subsection{Improvement by the energy deposit implementation\label{ssec:reconstruction:enedep}}

\begin{figure}[h]
    \centering
    \includegraphics[width = 0.65\textwidth]{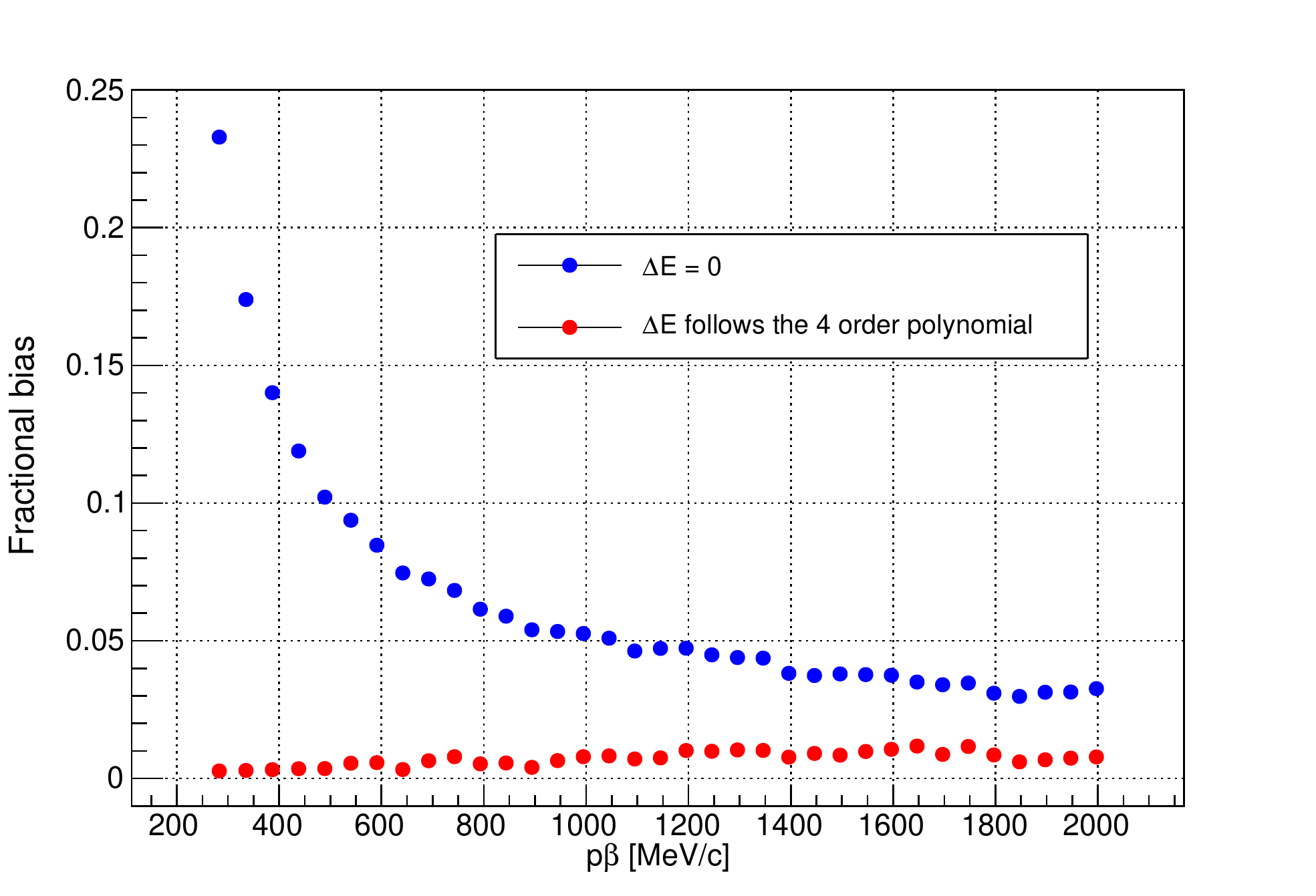}
    \includegraphics[width = 0.65\textwidth]{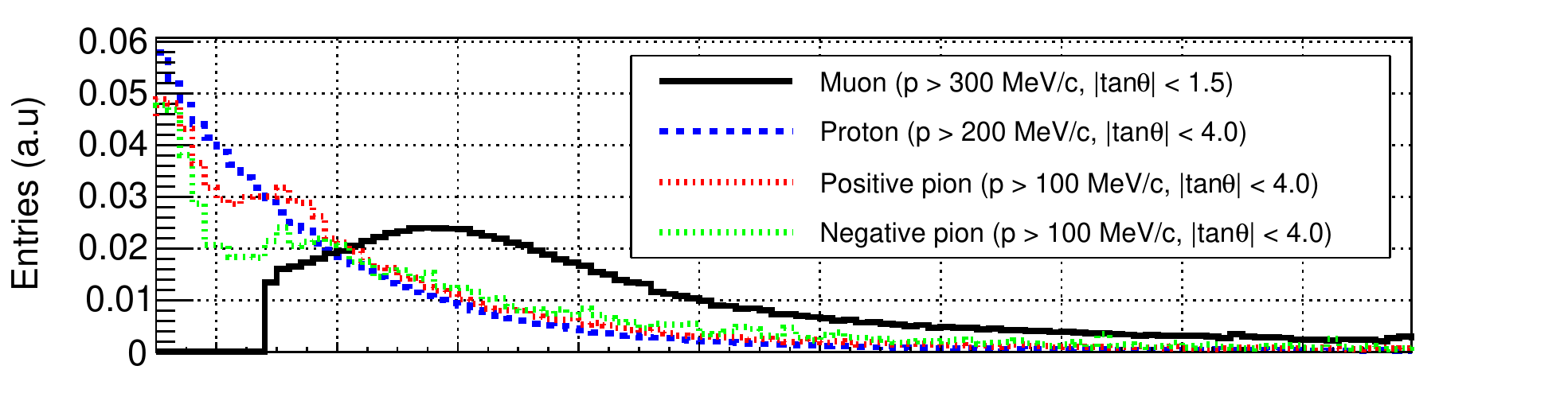}
    \caption{(Top) Fractional bias of the relative residual distribution of the $1/p\beta$ as a function of the momentum. The blue plots show the performance when $\Delta E = 0$, while the red ones consider the energy deposit. The vertical errors are negligible. (Bottom) Expected distributions of true $p\beta$ of the charged particles from the neutrino interactions in the NINJA experiment.\label{fig:enedep_compare_bias}}
\end{figure}

\begin{figure}[h]
    \centering
    \includegraphics[width = 0.65\textwidth]{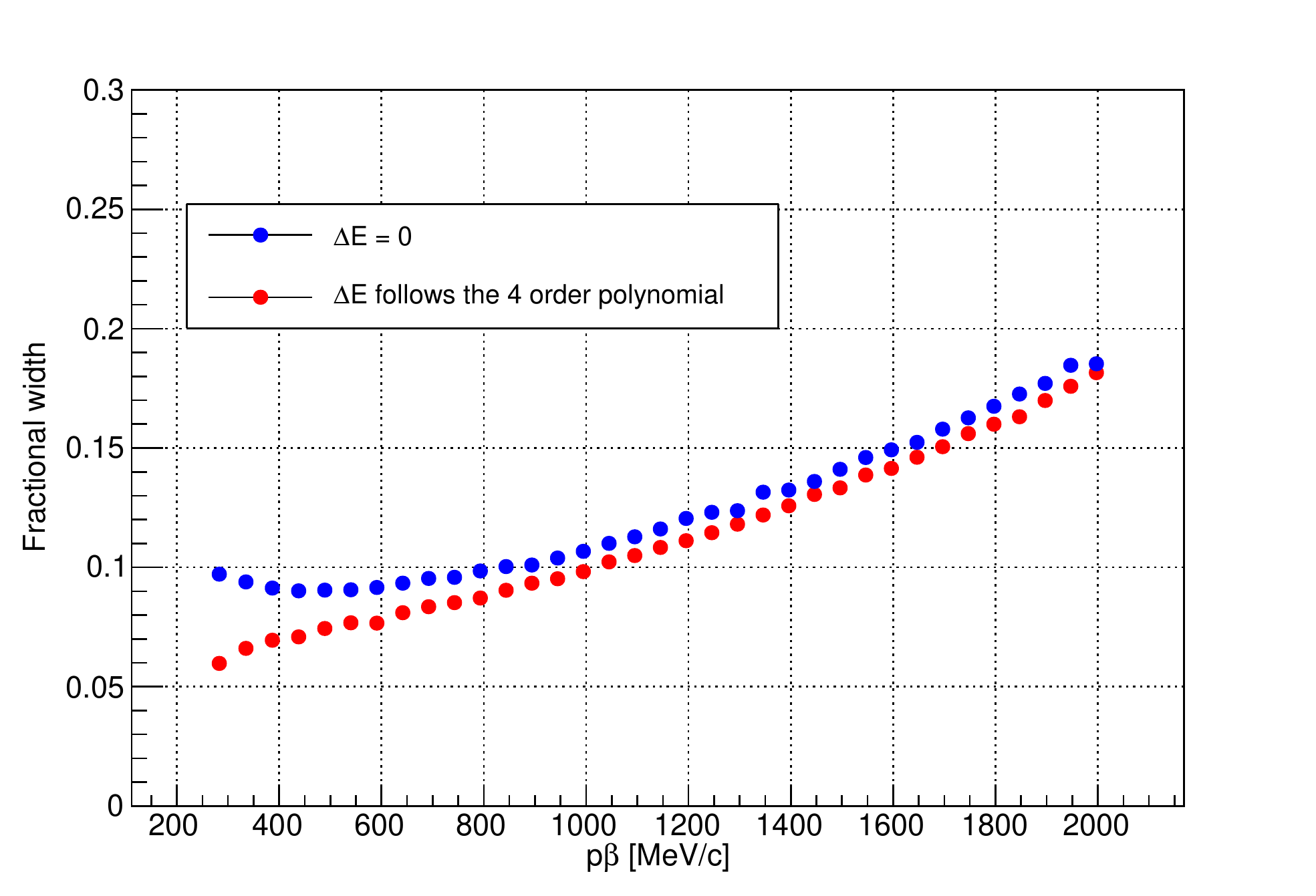}
    \includegraphics[width = 0.65\textwidth]{particle_distribution_com.pdf}
    \caption{(Top) Fractional width of the relative residual distribution of the $1/p\beta$ as a function of the momentum. The blue plots show the performance when $\Delta E = 0$, while the red ones consider the energy deposit. The vertical errors are negligible. (Bottom) Expected distributions of true $p\beta$ of the charged particles from the neutrino interactions in the NINJA experiment.\label{fig:enedep_compare_width}}
\end{figure}

In this new method, the implementation of the energy deposit plays an essential role to improve the performance of the previous method~\cite{Hiramoto:2020gup, Oshima:2020ozn, NINJA:2022zbi}.
This improvement is studied by comparing the results in Sect.~\ref{ssec:reconstruction:effect_precision} to the case without energy deposit implementation.
The reconstruction without energy deposit implementation is studied by setting $\Delta E_i = 0$ in Eq.~\eqref{eq:energy_deposition}, i.e. $p\beta$ is assumed to be unchanged as $(p\beta)_i =  (p\beta)_0$ along its trajectory.
The muon particle gun simulation from the most upstream water layer to the downstream perpendicularly to the film surface is also used in this study.
Figures~\ref{fig:enedep_compare_bias} and \ref{fig:enedep_compare_width} show the fractional bias and width of the relative residual distribution of the $1/p\beta$.
When the energy deposit is not considered, the bias is not negligible.
In particular, it is more than 10--20\% below $p\beta = \SI{500}{\MeV}/c$.
In a high-momentum region, the bias is expected to be reduced since the fraction of the energy deposit to the initial energy is small.
For $\SI{1}{\GeV}/c$ muon perpendicularly penetrating the ECC volume, the total energy deposit in the detector volume is expected to be $\SI{0.6}{\MeV}/\text{unit} \times 69\,\text{iron plates} \simeq \SI{40}{\MeV}$ from Fig.~\ref{fig:iron_bb_pol4_compare}.
When we consider the additional energy deposit by the water layers and other materials, it represents 8\% of the initial energy.
We can assume that the reconstructed momentum value without energy deposit is the average of the incident and final momenta of the particle in the detector, thus the 8\% energy deposit leads to 4\% fractional bias.
The value is almost consistent with Fig.~\ref{fig:enedep_compare_bias}.
This reduction of the fractional bias has a large importance in the analysis of the NINJA experiment.
The histograms in the bottom of Figs.~\ref{fig:enedep_compare_bias} and \ref{fig:enedep_compare_width} show the distributions of true $p\beta$ of the charged particles from the neutrino interactions in the water ECC with our typical detection acceptances.
These distributions are simulated by the NEUT neutrino event generator~\cite{Hayato:2002sd,Hayato:2009zz,Hayato:2021heg}.
A large portion of the particles have $p\beta$ below $\SI{1}{\GeV}/c$ with originally 5--25\% fractional bias.
The implementation of the energy deposit reduces such a bias to less than 1\% in the new method and highly improves the measurement accuracy.

The fractional width, which corresponds to the momentum resolution, is mainly determined by the number of the measurements of the scattering angles.
Thus, it is not largely changed, particularly in the high-momentum region.
In the low-momentum region, on the other hand, the difference is clearer.
In this region, the initial momentum is constrained by both $\sigma$ and the $\Delta E$ as $\beta$ changes.
Thus, the momentum resolution also improves when the energy deposit is considered.

\section{Validation by a momentum measurement with the track range\label{sec:range_validation}}

The validation of the method is done using data acquired in the NINJA experiment physics run (J-PARC E71a)~\cite{JparcE71Tdr:2019}.

\subsection{Detector setup in the NINJA experiment\label{ssec:range_validation:ninja_detectors}}

\begin{figure}[h]
    \centering
    \includegraphics[width = 0.7\textwidth]{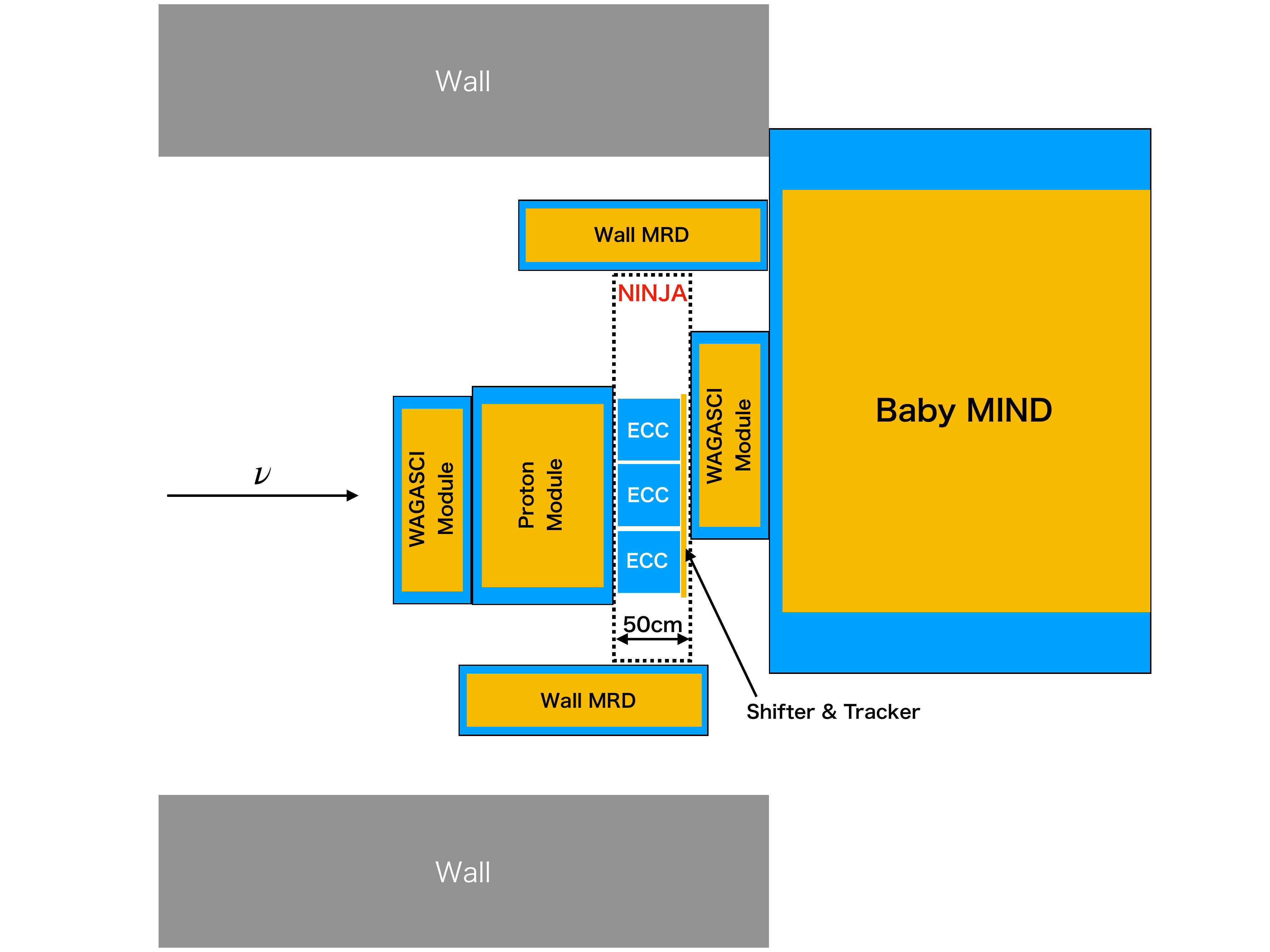}
    \caption[PRa detector setup]{Detector setup of the NINJA experiment physics run viewed from the top. The NINJA detectors are surrounded by the WAGASCI detectors in the T2K experiment. The blue and orange regions show the total and scintillator volumes of each detector, respectively. \label{fig:pra_detector_setup}}
\end{figure}

The detector environment of the NINJA experiment physics run is shown in Fig.~\ref{fig:pra_detector_setup}.
The water ECCs and other detectors are surrounded by the WAGASCI detectors in the T2K experiment in the J-PARC Neutrino Monitor building.
The NINJA experiment uses three kinds of detectors.
\begin{enumerate}
    \item Water ECC\\
    The structure of the water ECC is already described in Sect.~\ref{sec:introduction}.
    One chamber sizes around $\SI{30}{\cm} \times \SI{30}{\cm} \times \SI{30}{\cm}$ and nine ECCs are installed in the physics run.
    The chambers are $3 \times 3$ arranged on a plane perpendicular to the beam direction.
    \item Baby MIND~\cite{Yasutome:2022ems}\\
    Baby MIND is one of the T2K near detectors and is used as a muon range detector in the NINJA physics run.
    It consists of 18 scintillator planes and 33 magnetized iron planes.
    \item Timestamp detectors\\
    To connect muon tracks detected in the ECCs and Baby MIND, two kinds of timestamp detectors are installed.
    One is a scintillator tracker~\cite{Odagawa:2022crm} and the other is an emulsion shifter.
    The emulsion shifter has several emulsion films on walls moving different timing intervals.
    After the development of the emulsion films, the positional differences between each plane gives the timing information to each track~\cite{Oshima:2020ozn}.
    The timestamp detectors are installed as shown in Fig.~\ref{fig:pra_detector_setup} and match muon tracks between the ECCs and Baby MIND.
\end{enumerate}

\subsection{Event sample\label{ssec:range_validation:sample}}

The performance of the momentum reconstruction with MCS is evaluated using muon tracks coming from the upstream wall of the detector hall, penetrating the water ECC, and stopping inside the fiducial volume of Baby MIND.
To obtain such tracks, the following selections are applied to the data.
\begin{itemize}
    \item The track is required to be matched between Baby MIND and the water ECCs using the scintillation tracker~\cite{Odagawa:2022crm} and the emulsion shifter.
    \item The track is required to penetrate through the central water ECC from the most upstream emulsion film to the most downstream one of it.
    \item The track is required to be uniquely connected among the detectors.
    \item The track is required to stop inside the fiducial volume of Baby MIND to measure the momentum using the track range.
\end{itemize}

The data used in this validation correspond to the ones produced by $4.7\times 10^{20}$ protons on target in the J-PARC neutrino beamline.
Three-dimensional tracks in the WAGASCI detectors and the track matching between them are processed in the T2K experiment.
The three-dimensional tracks reconstructed in Baby MIND are extrapolated to the scintillation tracker and emulsion shifter, and the track matching among the detectors is done.
The details are described in Ref.~\cite{Odagawa:2022crm}.

\subsection{Comparison of the reconstruction with MCS and the track range\label{ssec:range_validation:performance}}

\begin{figure}[h]
    \centering
    \includegraphics[width = 0.7\textwidth]{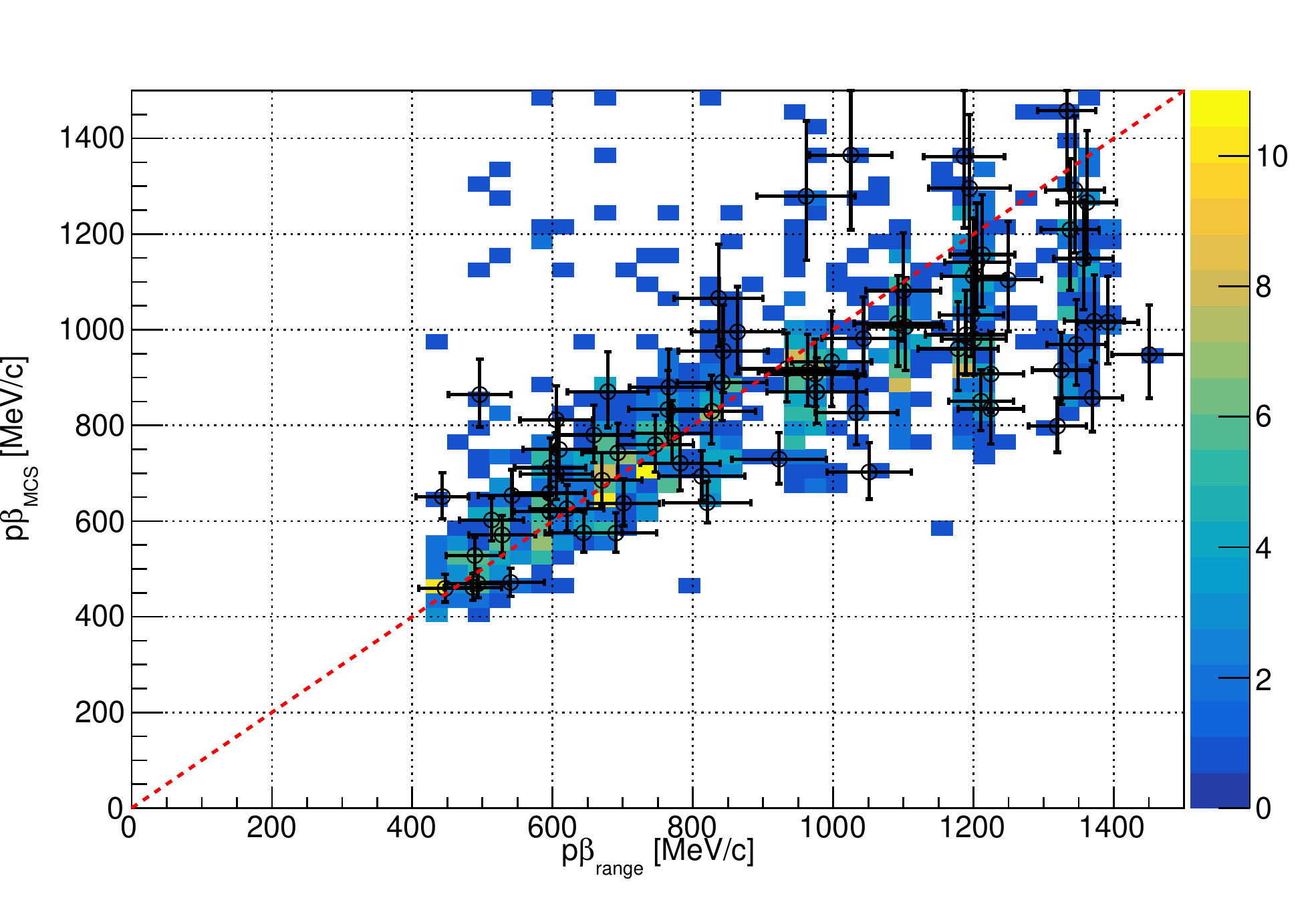}
    \caption{Two-dimensional correlation plot between the momentum reconstructed with MCS and that with the track range. The plots with measurement errors are 10\% randomly selected from the total entries. The distribution of $p\beta_\mathrm{range}$ represents the structure of the scintillator planes and quantized especially in high-momentum region.\label{fig:mcs_range_2d}}
\end{figure}

Baby MIND measures the path length of a charged particle in materials.
The momentum of the particle is reconstructed with the track range.
According to the studies in the T2K experiment, the momentum resolution of this method is approximately 10--15\% for the muons stopping inside the fiducial volume of Baby MIND, i.e. up to around $\SI{1.5}{\GeV}/c$.
Using the data sample described in Sect.~\ref{ssec:range_validation:sample}, the momentum of each track is reconstructed by two independent methods: MCS and the track range.
Figure~\ref{fig:mcs_range_2d} shows the two-dimensional correlation plot of the reconstructed momenta, $p\beta_\mathrm{MCS}$ and $p\beta_\mathrm{range}$.
Both momenta are reconstructed at the front of the ECC.
The reconstruction with the track range also considers the energy deposit in the ECC and the downstream WAGASCI detector.
The colored two-dimensional histogram shows the all samples, while 10\% of them are selected as plots with measurement errors.
The plots are selected completely randomly so as not to bias the result.
Some tracks show much higher value of $p\beta_\mathrm{MCS}$ than that of $p\beta_\mathrm{range}$.
This may be attributed to the wrong connection of tracks between Baby MIND and the ECC, or pion misidentification as muon in Baby MIND.
If a charged pion is misidentified as a muon, the track range will be small due to the hadron interaction with the materials in Baby MIND, while $p\beta_\mathrm{MCS}$ will show more correct reconstructed value.
Here, the momentum reconstructed with MCS shows saturation around $\SI{1}{\GeV}/c$ since the scattering angles are small and comparable to the angular resolution of the nuclear emulsion films.
In this momentum region, $\sigma_\mathrm{HL} = 2 \times 10^{-3}$, while $\epsilon_\mathrm{rad'} = 1.7 \times 10^{-3}$ for $\tan\theta \simeq 0$.
In the momentum reconstruction with MCS, 1.5\% of the scattering angles are discarded from the calculation of the likelihood, since they are larger than three times the RMS of them.

\begin{figure}[h]
    \centering
    \includegraphics[width = 0.67\textwidth]{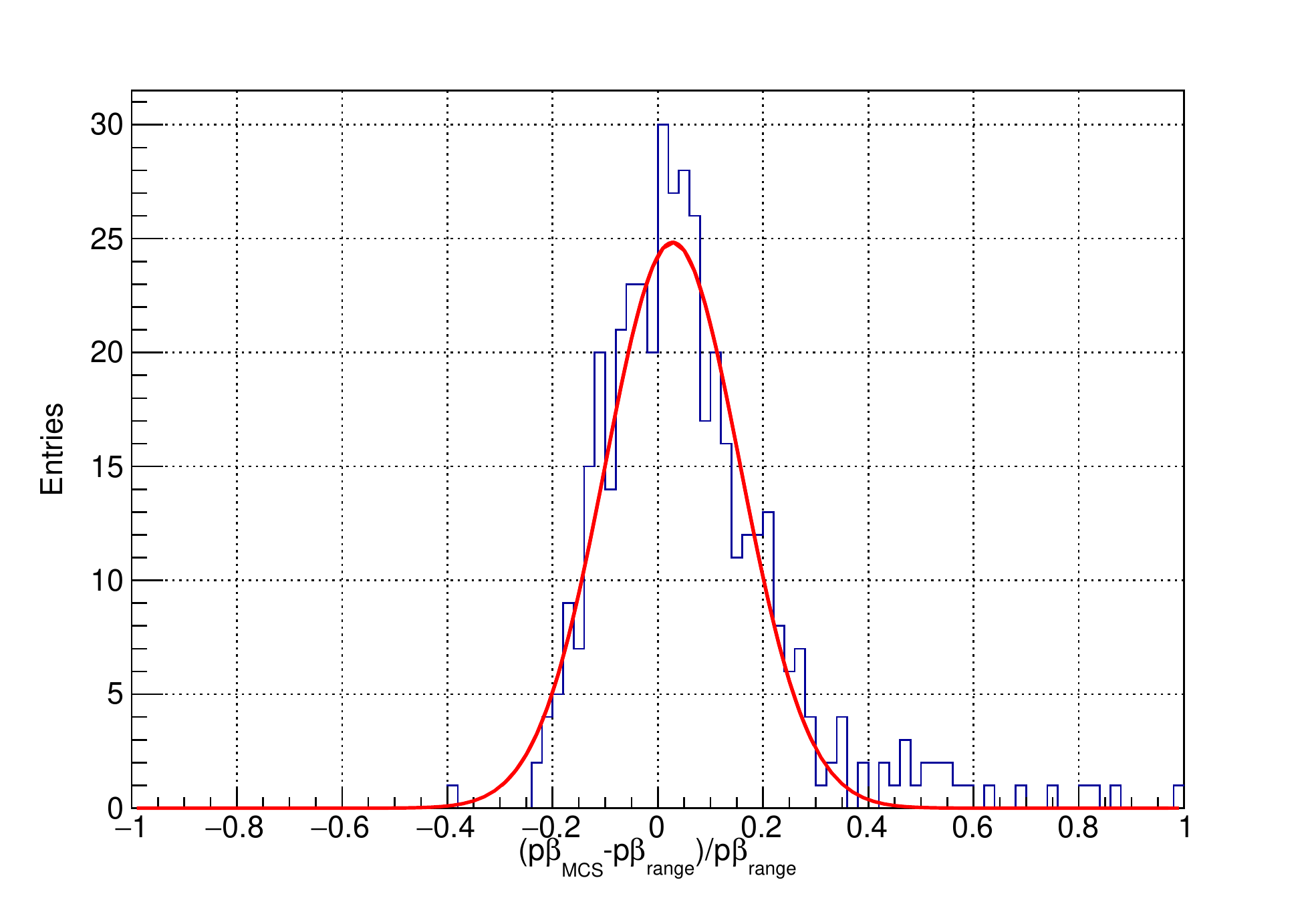}
    \caption{The residual distribution of $p\beta$ with $\SI{400}{\MeV}/c < p\beta_\mathrm{range} < \SI{900}{\MeV}/c$. The red line shows the result of a Gaussian fitting.\label{fig:pbeta_residual_w_fit}}
\end{figure}

The two methods are shown to be consistent between up to $\SI{1}{\GeV}/c$, within the experimental uncertainties, including the angular resolution of the nuclear emulsion films, the thickness of penetrating materials, or $S$ parameter tuning.
Figure~\ref{fig:pbeta_residual_w_fit} shows the residual distribution of $p\beta$ with $\SI{400}{\MeV}/c < p\beta_\mathrm{range} < \SI{900}{\MeV}/c$.
The mean and width of the fitted Gaussian function are 3\% and 13\%, respectively.
Including the systematic uncertainty of the fitting range, the mean is consistent with zero and the reconstructed values of the two methods agree.
The agreement of the two results experimentally verifies the new method of momentum reconstruction with MCS.
This result indicates that the new method of the momentum reconstruction with MCS will be useful in the NINJA experiment.

\section{Prospects\label{sec:prospects}}

The method described in this paper will be used for the momentum measurement of charged particles from the neutrino interactions in the NINJA experiment physics run.
In particular, not only muons stopping inside the fiducial volume of Baby MIND, but also ones escaping from side or penetrating through it, can be measured.
Therefore, the measurable phase space of muon momentum and angle will be extended from the case using only the momentum reconstruction with the track range.

Several possibilities to improve the method are presented as follows.
The momentum resolution is typically determined by the number of the measurements of the scattering angles.
The low-momentum particles, especially protons, stop inside the detector volume, which reduces the number of the measurements.
The number of the measurements can be increased by using the scatterings in the water layers in addition to those in the iron plates.
Although the particles are less scattered off by a \SI{2.3}{\mm}-thick water layer compared to a \SI{500}{\um}-thick iron plate, it will have an impact on low-momentum particles, which scatter largely enough compared to the angular resolutions.
If the scattering angles across each water layer are available in the method, the number of the measurements of the scattering angles used in the reconstruction will be almost doubled, thus the momentum resolution will be improved.

On the other hand, the saturation is the main problem in the high-momentum region.
It is caused by the fact that the magnitude of MCS is less than the angular resolution.
It may be able to be improved by calculating the scattering angle between a pair of films over more than one iron plate.
In this case, the number of the measurements of the scattering angles decreases but MCS shows larger angles, e.g. the magnitude of MCS across three iron plates is typically comparable to the angular resolution when $p\beta = \SI{1.9}{\GeV}/c$.
Therefore, the bias and relatively low resolution observed in the high-momentum region could be improved.
In addition, using a combination of the scattering angles between a pair of films over $N\,(N = 1, 2, 3, \dots)$ iron plates would give much more precise reconstruction.

The method is evaluated and validated by the muon sample, but is also applicable to hadrons.
In the hadron momentum reconstruction, $M$ in Eq.~\eqref{eq:pbeta_to_energy} is replaced by the mass of the hadron, whereas other calculations remain the same.
The minimization of the negative log likelihoods with different masses would make it possible to identify the particles using MCS.
According to Fig.~\ref{fig:enedep_compare_bias}, the effect from the energy deposit is clearly reflected in the likelihood, especially in low-momentum region.
The energy deposit is a function of $\beta$, whereas MCS is a function of $p\beta$.
Thus, the different values of $M$ result in different values of the likelihood and the best $M$ may be most likely selected from the smallest negative log likelihood value.

\section{Conclusions\label{sec:conclusions}}

The momentum measurement of charged particles is important for a better understanding of the neutrino\nobreakdash-nucleus interactions in the NINJA experiment.
The momentum reconstruction with MCS is a useful technique to measure the momentum of such particles.
A new maximum likelihood-based method for the reconstruction in the water ECC is described in this study.
The scattering angle and likelihood are considered to achieve a better performance than the previous method used in the NINJA experiment.
In the present method, the energy deposit of charged particles is considered to reduce the bias observed in the low-momentum region.
In addition, one parameter of the Highland formula is phenomenologically scaled to improve the performance.

The method is evaluated using the MC simulation of the muon particle gun.
According to the simulation, the momentum resolution is 10\% for muons of $p = \SI{500}{\MeV}/c$ passing through around 70 iron plates perpendicularly.
The method shows 10-20\% momentum resolution depending on the momentum and emission angle of the particle.
In particular, the implementation of the energy deposit in the likelihood significantly reduces the bias of the low-momentum particles from at most 25\% to 1\%.

Using the tuned Highland formula, the method is then validated by data taken in the NINJA physics run.
The momentum of the muon tracks coming from the upstream wall of the detector hall and penetrating the water ECC is reconstructed using two methods.
One method uses the track range measured by the downstream muon range detector and the other applies MCS in the water ECC.
The performance of the method proposed here is evaluated by comparing the results from the two methods.
The two reconstructed momenta agree well in the region of $\SI{400}{\MeV}/c < p\beta_\mathrm{range} < \SI{900}{\MeV}/c$.

The method presented in this study will extend the measurable phase space of muons and hadrons from the neutrino interactions in the NINJA experiment.
Furthermore, the measurement of low-momentum charged particles from the neutrino\nobreakdash-water interactions will be performed with less measurement error.

\section*{Acknowledgment}

We thank the J-PARC staff for their accelerator performance and the NINJA collaboration for data acquisition and fruitful discussions.
We also acknowledge the T2K neutrino beam and WAGASCI/Baby MIND groups for the stable neutrino beam and detector operations.
We would like to thank Editage (\url{www.editage.com}) for English language editing.

This work is supported by the JST-SENTAN Program from the Japan Science and Technology Agency, JST, and MEXT and JSPS KAKENHI Grant Numbers JP17H02888, JP18H03701, JP18H05535, JP18H05537, JP18H05541, JP20J15496, and JP20J20304.

\bibliographystyle{ptephy}
%\bibliography{main}

\appendix

\end{document}